\magnification=\magstep0
\baselineskip=12pt
\lineskip=0pt
\lineskiplimit=0pt
\parindent=15pt
\parskip 10truebp

\def\pp{\parshape 2 0truecm 15truecm 2truecm 13truecm}
\def\deg{^\circ}
\def\ltsima{$\; \buildrel < \over \sim \;$}
\def\simlt{\lower.5ex\hbox{\ltsima}}
\def\gtsima{$\; \buildrel > \over \sim \;$}
\def\simgt{\lower.5ex\hbox{\gtsima}}
\def\bfR{{\rm R}}
\def\bfS{{\rm S}}
\def\bfN{{\bf N}}
\def\bfr{{\bf r}}
\def\bfs{{\bf s}}
\def\bfrp{{\bf r}^\prime}
\def\bfk{{\bf k}}

\def\etal{{\it et al.~\/}}
\def\kms{\ifmmode {\rm \ km \ s^{-1}}\else $\rm km \ s^{-1}$\fi}
\def\mpc{$\ {h^{-1}\rm Mpc}$}
\def\bfv{{\bf v}}

\def\rr{{\scriptscriptstyle R}}
\def\bfSl{{ {\bf S}_{l} }}
\def\bfNl{{ {\bf N}_{l} }}
\def\np{{n^\prime}}
\def\npp{{{n^\prime}^\prime}}

\hfill\hfill\hfill Submitted MNRAS 12 July 1996

\hfill\hfill\hfill Corrected 5 August 1996 (AMW)

\noindent 
{\bf WIENER RECONSTRUCTION OF THE IRAS 1.2Jy GALAXY REDSHIFT SURVEY:} 

\noindent 
{\bf COSMOGRAPHICAL IMPLICATIONS}
\bigskip

\bigskip
Matthew Webster$^{1,3}$, Ofer Lahav$^1$ \& Karl Fisher$^{1,2}$
\bigskip

(1) Institute of Astronomy, Madingley Road, Cambridge CB3 0HA, UK

(2) Institute for Advanced Study, Olden Lane, Princeton, NJ 08540, USA

(3) email: mwebster@ast.cam.ac.uk

\bigskip
\noindent {\bf Abstract }
\medskip

We utilise the method of Wiener reconstruction with spherical
harmonics and Bessel functions to recover the density and velocity
fields from the IRAS 1.2Jy redshift survey. The reconstruction
relies on prior knowledge of the 
IRAS power-spectrum
and the combination of density and bias parameters
$\beta\equiv\Omega^{0.6}/b$. The results are robust to changes
in these prior parameters and the number of expansion
coefficients. Maps of these fields are presented in a variety of
projections. Many known structures are observed, including clear
confirmations of the clusters N1600 and A3627. The Perseus-Pisces
supercluster appears to extend out to $\sim 9000 \kms$, and the
reconstruction shows `backside infall' to the Centaurus/Great
Attractor region. A qualitative comparison of the reconstructed IRAS
gravity field with that from Tully-Fisher peculiar velocity
measurements (Mark III) shows reasonable agreement. The Wiener
reconstruction of the density field is also the optimal reconstruction
(in the minimum variance sense) of any quantity which is linear in the
density contrast. We show reconstructions of three such
quantities. 
The misalignment angle between the IRAS and CMB Local
Group dipoles is only $13\deg$ out to $5000 \kms$, but increases
to $25\deg$ out to $20,000 \kms$. 
The reconstructed IRAS bulk
flow out to $5000 \kms$ is $\sim 300 \kms$, which agrees in amplitude
with that derived from the Mark III peculiar velocities ($\sim 370
\kms$). 
However, the two bulk flow vectors deviate by $\sim 70\deg$. Finally,
moment of inertia analysis shows that the Wiener reconstructed
Supergalactic Plane is aligned  within $\sim 30 \deg$ of  that defined by de
Vaucouleurs.

\bigskip
\noindent {\bf 1. Introduction} 
\medskip

Observations of large scale structure have been advancing
spectacularly. Both in terms of redshifts and galaxy peculiar
velocities, the number of measurements has grown by more than a factor
of ten over the past decade. Hand in hand with observational progress,
a plethora of statistical methods has been developed to analyse the
data, each attempting to extract as much astrophysical information as
possible. Accurate determination of the galaxy distribution and
velocity field would not only place constraints on cosmological
parameters, but would also provide an insight into the mechanisms of
structure formation which generated the complex pattern of sheets and
filaments we observe today.

Redshifts have been determined for many more galaxies than those for
which we have direct distance measurements. Over the next few years,
automated surveys such as the 2-degree-Field (2dF) and Sloan Digital
Sky Survey (SDSS) will increase this lead even further, as over a
million new galaxy redshifts become available. This wealth of data has
inspired a great deal of work on techniques for reconstructing the
velocity field on the basis of redshifts alone. Given the assumptions
of linear mass-to-light biasing, and purely linear structure
evolution, the mapping between real and redshift-space is well defined
(Kaiser 1987). However, in applying reconstruction methods there is
freedom of choice regarding (i) the functional representation
(e.g. Cartesian, Fourier, Spherical Harmonics, or Wavelets) and (ii)
the filtering, or smoothing scheme (e.g. a Gaussian sphere, a sharp
cut-off in Fourier space, or a Wiener filter). Smoothing is necessary
to reduce sampling noise before the real-space density field can be
calculated from that in redshift-space.

The first reconstruction technique to be applied to redshift surveys
was based on iterative solution of the equations of linear theory,
pioneered by Yahil \etal (1991), hereafter YSDH (see also Yahil 1988;
Strauss \& Davis 1988). This involves solving for the gravity field in
redshift space, which can then be used to derive an estimate of the
peculiar velocities for a given value of $\beta \equiv
\Omega^{0.6}/b$. These velocities allow the redshifts to be corrected,
providing an updated set of distance estimates. This is repeated until 
the distance estimates converge. YSDH use variable smoothing, but their
smoothing regime is not rigorously formulated. Variants of this
technique have been successfully applied to IRAS selected galaxy
catalogues (YSDH; Kaiser \etal 1991).

The reconstruction procedure used in this paper follows Fisher \etal
(1995$a$; hereafter FLHLZ) and is {\it non-}iterative. The density field
in redshift-space is expanded in terms of spherical harmonics and
Bessel functions. The real-space density, velocity and potential
fields are reconstructed using linear theory and a Wiener filter which
assumes a given power spectrum and noise level. It is important to 
note that, as opposed to ad-hoc smoothing schemes, the smoothing
due to a Wiener 
filter is determined by the sparseness of data
relative to the expected signal.

Cosmography using spherical harmonics is not a new technique (Peebles
1973), but it has enjoyed a renaissance with the advent of near whole-sky
galaxy surveys (e.g. Fabbri \& Natale 1989; Regos \& Szalay 1989;
Lynden-Bell 1991; Scharf \etal 1992;  Scharf \& Lahav 1993; 
Fisher, Scharf \& Lahav 1994, hereafter FSL; FLHLZ; Lahav 1994; 
Nusser \& Davis 1994; Heavens \& Taylor 1995). 
Similarly, the Wiener filter (Wiener 1949; hereafter WF)
has long been used in engineering to recover the best estimate of the
true signal from one corrupted by imperfect measurement. Recently,
Wiener filtering has been applied to a number of cosmological
reconstruction problems. Lahav \etal (1994; hereafter LFHSZ)
reconstructed the angular distribution of IRAS galaxies, while Bunn \etal
(1994) applied the method to the temperature fluctuations in the {\it
COBE} DMR maps. Kaiser \& Stebbins (1991) used a similar
method to reconstruct velocity fields. Reviews of the WF and linear
estimation methods can be found in Press \etal (1992), Rybicki \&
Press (1992) and Zaroubi \etal (1995; hereafter ZHFL).

FLHLZ applied the 3-dimensional Wiener reconstruction to the
IRAS 1.2Jy survey, showing preliminary predictions for the density, 
velocity and potential fields. 
In this paper, we expand on the initial results in
FLHLZ, presenting detailed maps of the reconstructed 
fields, as well as optimal determinations 
of the Local Group dipole,
bulk flows and the extent of the Supergalactic Plane.
We compare the IRAS reconstructed peculiar velocity
field to that derived from Tully-Fisher distances (Mark III) and examine 
the sensitivity of the WF method to choice of parameters.

In \S 2, we give a brief summary of the decomposition and filtering
techniques used in this reconstruction and the data set to which they
have been applied. \S 3 discusses the role of the prior model in the
WF and the methodology behind selection of parameters. Results of the
reconstruction are shown in \S 4, and these are compared to Mark III
data in \S 5. As detailed below, the choice of prior model is central
to the operation of the WF and \S 6 discusses the qualitative
influence of these parameters on the reconstructed density and
velocity fields. In \S 7 and \S 8, we show reconstructions of the
Local Group dipole and bulk flow respectively; both determined 
directly from harmonic coefficients. Similarly, a reconstruction of
the Supergalactic Plane is shown in \S 9, and its alignment compared
to previous measurements. Finally \S 10 contains a discussion of
the work in this paper.

\bigskip
\noindent {\bf 2. Wiener reconstruction of IRAS data}
\medskip

\noindent {\bf 2.1 Reconstruction Method}

This paper concentrates on cosmography, presenting results from the
application of the WF method to IRAS 1.2Jy survey data. Hence, we
include here only a brief introduction to our reconstruction technique. 
The approach we use is discussed in detail by FLHLZ, while the formalism
of Wiener filtering as it pertains to LSS reconstruction can be found
in ZHFL. 

A density field, $\rho$, can be expanded in terms of spherical
harmonics, $Y_{lm}$, and Bessel functions, $j_l$:
$$
\rho(\bfs) = 
\sum\limits_{l=0}^{l_{\rm max}}
\sum\limits_{m=-l}^{+l}
\sum\limits_{n=1}^{n_{\rm max}(l)}\, 
C_{ln}\, \rho_{lmn}\, j_l(k_n s)\, Y_{lm}(\hat\bfs)
\qquad .
\eqno (1)
$$
The discrete $k_n$'s are chosen according to the boundary conditions,
so as to make the set orthogonal (see Appendix A in FLHLZ). This
process is analogous to Fourier decomposition, but instead using a set
of spherical basis functions. The data from a redshift catalogue can
be seen as a set of $N$ discrete points, $\bfs_i$, each giving the
direction and redshift of a galaxy. These are used to estimate the
underlying density field in redshift space, $\hat\rho^\bfS(\bfs)$,
expanded as in equation 1. Here, $C_{ln}$ are normalisation constants,
while the harmonic coefficients are given by
$$
\hat\rho_{lmn}^{\bfS} = 
\sum\limits_{i=1}^N {{1}\over{\phi(s_i)}}
j_l(k_ns_i) Y^\ast_{lm}(\bf {\hat\bfs_i} ) \qquad ,
\eqno (2)
$$
where $\phi(s_i)$ is the spherical selection function of the survey,
evaluated at the radius of the $i^{th}$ galaxy. From $\hat\rho(\bfs)$,
we can derive an estimate of the overdensity field, $\delta(\bfs) =
(\rho(\bfs) / {\bar\rho}) - 1$, in terms of the same basis functions,
with coefficients $\hat\delta_{lmn}^{\bfS}$. This is not only more
useful for dynamics, but also has, by construction, a mean of zero,
which is useful later on in the reconstruction method.

As redshifts are line-of-sight measurements, peculiar velocities will
introduce only radial distortion in redshift space. Given that
$\delta^\bfS$ is expressed in terms of orthogonal radial and angular
components, redshift distortion can be seen as coupling between the
radial modes in real and redshift space, leaving the angular modes 
unaffected. This can be expressed by introducing a matrix which links 
radial modes in the density fluctuation fields,
$$
\hat\delta_{lmn}^\bfS = \sum\limits_{n^{\prime}} \,
({\bf Z}_l)_{nn^{\prime}}\, \hat\delta_{lm{n^\prime}}^\bfR
\qquad ,
\eqno (3)
$$
where ${\bf Z}_l$ depends on the chosen value of $\beta$ and the
selection function for the survey (see Appendix D in FLHLZ). Here we
use superscripts R and S to denote real and redshift space
respectively. Note that this is only applicable for all-sky surveys;
in samples with incomplete sky coverage, there will also be more
complicated coupling of angular modes.

In a perfect galaxy catalogue, with arbitrarily high sampling, the
real-space harmonics could be estimated by simple inversion of the
coupling matrix in equation 3. However, with real data,
straightforward inversion is often unstable, and a regularisation
scheme is required. Without such a scheme, shot noise from the
discreteness of the galaxy distribution might be amplified by
inversion, leading to an estimate that was far from
optimal. Application of a Wiener filter to the inversion process
allows one to reconstruct the minimum variance solution, assuming a
prior power spectrum and a shot noise level. Appendix A provides a
summary of the WF method. For further details, please refer to Rybicki
\& Press (1992), LFHSZ and ZHFL.

In our problem, the redshift coupling matrix represents the response
function of the system, acting on the real-space distribution plus
shot noise to give the observed, $\delta^\bfS$. Hence, the WF
estimation of the real space overdensity field is
$$
{(\hat\delta_{lmn}^{\bfR})}_{\rm{wf}} = \sum\limits_{\np\npp}
\left( \bfS_l \left[ \bfS_l + \bfN_l \right]^{-1}\right)_{n\np} 
\left( {\bf Z}_l^{-1}\right)_{\np\npp} \hat\delta_{lm\npp}^\bfS
\qquad .
\eqno (4)
$$
The assumed true signal matrix, $\bfS$, depends on the power spectrum,
while the noise matrix, $\bfN$, is a function of the mean density and
selection function (see LFHSZ). Incidentally, as Rybicki \& Press
(1992) point out, the WF is in general a biased estimator of the mean
field unless the field has zero mean. As mentioned above, the density
fluctuation field has zero mean.

Just as the density field can be decomposed into harmonics, so too can
the radial and transverse components of the velocity field, and the
potential field. In the case of radial velocities, linear theory
allows the velocity due to inhomogeneities within $r < R$ to be
expressed directly in terms of the harmonics of the overdensity
field. Again, this can be expressed by introducing a coupling matrix,
in a similar manner to that in equation 3, such that
$$
{(v_{lmn})}_{\rm{wf}} =\beta \sum\limits_{\np} \left( {\bf
\Xi}_l\right)_{n\np} (\delta_{lm\np}^\bfR)_{\rm{wf}} 
\qquad .
\eqno (5)
$$
Transverse velocity and potential harmonics are also related to those
of the density field (Appendix C in FLHLZ).

To summarise, the redshift-space density field is estimated in terms
of orthogonal functions which are 
eigenfunctions of the Laplacian operator. For a given
choice of $\beta$, the radial coupling matrix linking real- and
redshift-space harmonics can be determined. Regularisation by a Wiener
filter, with a given power spectrum as its prior, allows inversion of
the coupling matrix and determination of the real-space density
field. Linear theory then allows the velocity and potential fields to
be determined from the estimated density field. We emphasise that the
method is {\it non}-iterative, and uses a filter which is optimal in
terms of assumed signal and noise characteristics. It provides a 
non-parametric and
minimum variance estimates of the density, velocity, and potential
fields which are related by simple linear transformations.

\bigskip
\noindent {\bf 2.2 The IRAS 1.2 Jy Survey} 
\medskip

The IRAS surveys are uniform and complete down to Galactic latitudes
as low as $\pm 5\deg$ from the Galactic plane. This makes them ideal
for estimating whole-sky density and velocity fields. Here, we use the
1.2 Jy IRAS survey (Fisher \etal 1995$b$), consisting of 5313 galaxies,
covering 87.6\% of sky with the incomplete regions being dominated by
the 8.7\% of the sky with $|b|<5\deg$. 

In principle, the method can be extended to explicitly account for the
incomplete sky coverage. We adopt the simpler approach of smoothly
interpolating the redshift distribution over the missing areas using
the method described in YSDH. Given this, we can assume a purely radial 
selection function. In LFHSZ, we examined the
validity of this interpolation and found its effect on the computed
harmonics to be negligible for $l\simlt 15$ for the geometry of the
1.2 Jy survey. We therefore use the
interpolated catalogue (and a simplified formalism) for our reconstructions.

\bigskip
\noindent {\bf 3. Selection of a prior model} 
\medskip

% {\it Figure 1 - Relationship of RMS scatter to $\beta$}

Throughout this paper, the density field is expanded
within a spherical volume of radius $R = 20,000\kms$. Boundary
conditions are imposed at the edge of this volume, forcing the radial
modes to be discrete rather than continuous. We have demanded that the
logarithmic derivative of the gravitational potential be continuous at
the boundary; a discussion of various boundary conditions is contained
in Appendix A of FLHLZ. We then truncate the summation over angular
and radial modes at some $l_{\rm max} = 15$ and $k_nR<100$ respectively,
effectively determining the resolution of the density field (please
see Appendix B of FLHLZ for further details). The qualitative effects
of $l_{\rm max}$ on resolution are discussed in \S 6. 

Beyond the definition of the expansion, the reconstruction method
itself has a number of free parameters. The WF assumes a prior power
spectrum, while the redshift coupling matrix depends on choice of
$\beta$. Fortunately, the shape and amplitude of the power spectrum
has been relatively well determined for IRAS galaxies. The power
spectrum is described on scales $\simlt 200$\mpc by a CDM power
spectrum with shape parameter $\Gamma$ (see Efstathiou, Bond, \& White
1992) in the range $\Gamma\simeq 0.2-0.3$ (Fisher \etal 1993; Feldman,
Kaiser, \& Peacock 1994). The normalisation of the power spectrum is
conventionally specified by the variance of the galaxy counts in
spheres of 8\mpc, $\sigma_8$. Fisher \etal (1994$a$) used the
projection of the redshift space correlation function to deduce the
value, $\sigma_8=0.69\pm0.04$ in {\it real} space. We have adopted the
WF prior given by $\Gamma=0.2$ and $\sigma_8=0.7$.

For simplicity, we assume linear, scale-independent biasing, where $b$
measures the ratio between fluctuations in the galaxy and mass
distribution, 
$$
(\delta\rho/\rho)_{\rm gal} = b(\delta\rho/\rho)_{\rm mass}
\qquad .
\eqno (6)
$$
In linear theory, the density and velocity fields are related by
$\beta\equiv\Omega^{0.6}/b$, where $\Omega$ is the density
parameter. The correct value of $\beta$ for IRAS is uncertain. Strauss
\& Willick (1995) review work on determining $\beta_{\rm IRAS}$ by a
variety of methods, giving values in the range $0.35 \leq \beta_{\rm IRAS}
\leq 1.28$. 

As discussed in FLHLZ, our reconstruction technique is essentially
perturbative, involving a Taylor expansion to first order in
$\Delta\bfv = ( \bfv(\bfr) - \bfv({\bf 0}))$, where $\bfv({\bf 0})$ is
the motion of the observer at the same frame as $\bfv(\bfr)$. Given
that the choice of calculation frame is arbitrary, it seems sensible
to select a frame in which $\Delta\bfv$ is minimal. It is known that
nearby galaxies (nearer than $\sim 2000\kms$) have small peculiar
velocities relative to the Local Group (LG), while very distant
galaxies are, on average, most likely to be at rest with respect to
the Cosmic Microwave Background (CMB). Rather than taking an ad-hoc
value for $\beta$, it makes sense to try the reconstruction in both
frames, and take the value of $\beta$ for which the two calculations
are in best agreement. 

We compare the two reconstructed line-of-sight peculiar velocities at
the position of all catalogued IRAS galaxies by calculating the RMS
scatter as function of $\beta$. Figure 1 shows that the minimum
scatter occurs at $\beta \approx 0.7$; this result is robust to changes
in the limiting radius of the reconstruction. There are many other possible
methods for selecting the prior value of $\beta$. For example, we can
choose the value which best reconstructs the motion of the LG with
respect to the CMB; this gives $\beta \approx 0.6$ to recover the 600
\kms. Another alternative is to find the $\beta$ for which the Local
Group motion is identical in both calculation frame, yielding $\beta
\approx 0.8$. FLHLZ used $\beta=1.0$, based on maximum likelihood
analysis in FSL. Hereafter we adopt $\beta=0.7$ based on
figure 1, which is well within the range 
quoted by various studies (Strauss \& Willick 1995).
In \S 6 we investigate the effects on the reconstruction of
changing $\beta$, $\Gamma$ and $\sigma_8$.

\bigskip
\noindent {\bf 4. Density and Velocity Maps}
\medskip

% {\it Figure 2 - Aitoff Diagrams showing D at 2000,4000,6000,8000 km/s}

% {\it Figure 3 - Aitoff Diagrams showing V at 2000,4000,6000,8000 km/s}

% {\it Figure 4 - Panel of 9 plots showing D at SGX, Y, Z =
% +4000,0,-4000 km/s}

% {\it Figure 5 - Panel of 9 plots showing D at SGX, Y, Z =
% +4000,0,-4000 km/s}

The maps shown in this section were derived by the technique detailed
above. In summary, the reconstruction was carried out within a sphere
of radius $R = 20,000 \kms$, with angular modes limited to $l_{\rm max} =
15$ and radial modes to $k_nR<100$, with a total of 6905
coefficients. A canonical set of values were adopted for cosmological
parameters, such that $\beta = 0.7$, $\Gamma = 0.2$ and $\sigma_8 =
0.7$. The reconstruction has a variable smoothing scale similar to a
Gaussian window of width, 
$\sigma_s$, proportional to the mean inter-particle separation 
(e.g. $\sigma_s = 436$ \kms\ at $r=4000$ \kms, $\sigma_s=626$ \kms 
at $6000$ \kms, and $\sigma_s=1130$ \kms at 10,000 \kms). 
The density contrast values shown in the diagrams and quoted below 
have been obtained under this smoothing regime.
Figures 2 
and 3 show Aitoff projections of the real-space density and radial 
velocity fields respectively, plotted in Galactic coordinates, 
evaluated across shells at various distances. Many previous papers have 
examined parts of the volume shown in these maps, identifying clusters 
and voids (e.g. Tully 1987, Pellegrini \etal 1990, Saunders \etal 1991). 
Where obvious analogues exist, structures are labelled in 
accordance with previous papers.

Figure 2a shows a density shell at $r = 2000\kms$. All major features have
already been identified, 
and there is the expected concentration of clusters in the
northern Galactic hemisphere. In particular, note
Fornax-Doradus-Eridanus ($l \simeq 180\deg$, $b \simeq -60\deg$),
N5846 ($l \simeq 350\deg$, $b \simeq 45\deg$), Virgo ($l \simeq
290\deg$, $b \simeq 70\deg$) and Ursa Major ($l \simeq 135\deg$, $b
\simeq 30\deg$). The foregrounds of Hydra ($l \simeq 270\deg$, $b
\simeq 10\deg$), and Centaurus ($l \simeq 310\deg$, $b \simeq 20\deg$)
are visible, and these structures can be seen to extend onto the next
shell. The observed shoulder on Hydra at $l \simeq 240\deg$, $b \simeq
0\deg$ is due to Puppis (e.g. Lahav \etal 1993), 
which appears as a distinct cluster at $r = 2500\kms$. The
overdensity marked C$\alpha$ ($l \simeq 10\deg$, $b \simeq -60\deg$) has
previously been identified as part of the Pavonis-Indus-Telescopium
(P-I-T) supercluster (Santiago \etal 1995). However, in this
reconstruction, C$\alpha$ appears to be quite distinct, peaking at $r \simeq 2500
\kms$ with $\delta_{\rm max} = 2.8$ on the smoothing scale 
described above. The Local Void (LV), as discussed by
Tully (1987) is clearly seen in the generally under-dense
region roughly spanning $330\deg < l < 120\deg$, $-50\deg < b <
30\deg$. Two additional voids are also marked, V$\alpha$ ($150\deg < l <
220\deg$, $-30\deg < b < 30\deg$) and V$\beta$ ($240\deg < l < 300\deg$,
$-90\deg < b < -20\deg$). The LV and V$\beta$ regions extend out to $r \simeq
4500 \kms$, while V$\alpha$ ends at $r \simeq 3500 \kms$.

Figure 2b is density at $r = 4000\kms$. Continuing from the previous shell,
the Hydra ($l \simeq 270\deg$, $b \simeq 25\deg$) and Centaurus/Great
Attractor (GA; $l \simeq 315\deg$, $b \simeq 15\deg$) superclusters
are shown clearly, both extending across $-10\deg < b < 45\deg$. The
foregrounds of P-I-T ($l \simeq 340\deg$, $b \simeq -25\deg$) and
Perseus-Pisces (P-P; $l \simeq 150\deg$, $b \simeq -15\deg$) can be
seen, as can Leo ($l \simeq 240\deg$, $b \simeq 75\deg$), Cancer ($l
\simeq 190\deg$, $b \simeq 25\deg$) and Cetus ($l \simeq 180\deg$, $b
\simeq -60\deg$). There is also confirmation of N1600 ($l \simeq 200$,
$b \simeq -28$), which is the strongest overdensity on this shell
($\delta_{\rm max} = 4.2$). The position of N1600 closely matches that
previously determined in both the IRAS 
(Saunders \etal 1991) and ORS (Santiago \etal 1995) surveys. 
Camelopardalis ($l \simeq 145\deg$, $b \simeq
30\deg$) is visible in the northern foreground of the Perseus-Pisces
supercluster (centre at $l \simeq 150\deg$, $b \simeq -15\deg$). The
two are linked by a cluster marked C$\beta$ ($l \simeq 135\deg$, $b \simeq
10\deg$), making a wall of continuous overdensity extending over
$-85\deg < b < 45\deg$. Extensions of LV and V$\alpha$ are also seen,
although the underdensity is considerably less smooth than at 
$r = 2000\kms$. Three dense regions on 
the shell do not correspond well to any
previous labelled clusters. C$\gamma$ ($l \simeq 110$, $b \simeq -70$) is the
foreground of a structure which extends out to $r \simeq 8000\kms$,
including the previously-identified A194 galaxy cluster at $\sim 6000
\kms$. C$\delta$ ($l \simeq 105$, $b \simeq -40$) is the foreground of a
cluster which extends to $r \simeq 6500 \kms$, where it merges into
Pegasus. Finally, C$\epsilon$ ($l \simeq 65$, $b \simeq 70$) is the foreground
of a weakly overdense wall that merges into the Great Wall structure
at $r \simeq 8000\kms$. 

Figure 2c is density at $r = 6000\kms$. Continuing from the previous shell,
note clear extensions of P-P ($l \simeq 150\deg$, $b \simeq -15\deg$),
P-I-T ($l \simeq 0\deg$, $b \simeq -30\deg$), Hydra ($l \simeq
280\deg$, $b \simeq 10\deg$), Leo ($l \simeq 270\deg$, $b \simeq
60\deg$), C$\gamma$ ($l \simeq 120\deg$, $b \simeq -70\deg$) and C$\epsilon$ ($l
\simeq 0\deg$, $b \simeq 70\deg$). Faint backgrounds can also be seen
for Cancer ($l \simeq 190\deg$, $b \simeq 30\deg$) and C$\delta$ ($l \simeq
110\deg$, $b \simeq -45\deg$). Orion ($l \simeq 190\deg$, $b \simeq
0\deg$), as shown on this shell, is the background of a strong cluster
centred at $r \simeq 5000\kms$, where it links Cancer and N1600 to form a
structure extending over $-45\deg < b < 45\deg$. A569 ($l \simeq
165\deg$, $b \simeq 20\deg$) is centred on this shell, and corresponds
very well to the position, extent and redshift (Abell et
al. 1989). The foreground of the Pegasus ($l \simeq 90\deg$, $b \simeq
-20\deg$) supercluster is seen weakly; in this reconstruction, Pegasus
extends to $r > 12,000 \kms$. There is also confirmation of the cluster
reported by Kraan-Korteweg \etal (1996), marked A3627 ($l \simeq
120\deg$, $b \simeq -8\deg$), which can be seen distinctly now that
Centaurus no longer swamps the region. The remaining cluster, labelled
C$\zeta$ ($l \simeq 40\deg$, $b \simeq -20\deg$) does not correspond well to
any known labels; it is the foreground of a large structure extending
to $r > 10,000\kms$, and appears to be quite separate from the nearby
P-I-T supercluster.

Figure 2d is density at $r = 8000\kms$. Continuing from the previous shell,
P-P ($l \simeq 150\deg$, $b \simeq 0\deg$), Leo ($l \simeq 260\deg$,
$b \simeq 45\deg$), A3627 ($l \simeq 320\deg$, $b \simeq -8\deg$) and
C$\zeta$ ($l \simeq 30\deg$, $b \simeq -35\deg$) extend out to 
$r \simeq 8000\kms$. The cluster marked 
C$\epsilon$ has merged into the body of the Great Wall
(GW; ($l \simeq 0\deg$, $0\deg < b < 90\deg$), which itself merges
with C$\zeta$ at $r \simeq 10,000\kms$ to form a wall extending over $-60\deg <
b < 90\deg$. Another wall is formed by Leo and Coma ($60\deg < l <
240\deg$, $b \simeq 75\deg$); this also seems to extend down beyond
the Galactic equator at greater distances. A779 ($l \simeq 170\deg$,
$b \simeq 45\deg$), A539 ($l \simeq 190\deg$, $b \simeq -20\deg$) and
A400 ($l \simeq 170\deg$, $b \simeq -40\deg$) all correspond to
clusters in the Abell catalogue. Cygnus ($l \simeq 80\deg$, $b \simeq
15\deg$) and C$\eta$ ($l \simeq 250\deg$, $b \simeq -5\deg$) are both
foregrounds of structures extending to $r \simeq 10,000\kms$. Notably, in
this reconstruction, P-P extends out to roughly 9000 \kms; far
further than usually thought. Out at $r > 6000\kms$, the peak
overdensity for P-P lies very close to the Galactic equator, and well
within the zone of avoidance (ZOA). It is important to remember that
C$\eta$ and P-P both lie largely within the $|b| < 5\deg$ region in which
data has been interpolated to fill the zone of avoidance. However,
both are strong overdensities on this shell ($\delta_{\rm P-P} \simeq
4$), and the shoulders are clearly visible in the IRAS data. 

Figure 3a shows line-of-sight velocity at $r = 2000 \kms$. The major feature
in this plot is the velocity dipole due to the Centaurus/GA region
(towards $l \simeq 300\deg$, $b \simeq 15\deg$). Also visible are the
effects of the P-P/Cancer/N1600 mass concentration, causing outflow
towards $l \simeq 165\deg$, $b \simeq -10\deg$. In figure 3b, at 
$r = 4000\kms$, some backside infall is visible in the Centaurus/GA region. The
strongest features are outflow towards P-P ($l \simeq 145\deg$, $b
\simeq -20\deg$), N1600 ($l \simeq 195\deg$, $b \simeq -20\deg$) and
Cancer ($l \simeq 190\deg$, $b \simeq 20\deg$), while weaker outflow
can also be seem towards the centre of P-I-T ($l \simeq 340\deg$, $b
\simeq -20\deg$). Figure 3c, at $r = 6000\kms$ is dominated by motion from 
the south Galactic pole towards the North Galactic pole. This is towards
the Great Wall in the northern Galactic hemisphere, while outflow
towards the A569/P-P region ($l \simeq 165\deg$, $-50\deg < b <
30\deg$) and the overdensity around C$\eta$ ($l \simeq 240\deg$, $b
\simeq -15\deg$) can also be seen. Backside infall is now seen behind P-I-T and
southern Centaurus. Finally, in figure 3d, at $r = 8000\kms$, we can see
evidence of the mass distribution at larger radii. Beyond 10,000\kms,
this reconstruction places large superclusters at $l \simeq 60\deg$,
$b \simeq 60\deg$ (forming out of the Great Wall), and $l \simeq
220\deg$, $b \simeq -30\deg$ (forming behind C$\eta$ and A400). These cause
the observed outflow, while strong infall can be seen behind Pegasus
($l \simeq 90\deg$, $b \simeq -25\deg$) and Hydra ($l \simeq 285\deg$,
$b \simeq 5\deg$).

Figure 4 shows the density field on planes at {\it Supergalactic} $x$,
$y$, $z$ = $0$, $\pm4000$ \kms, while figure 5 shows the corresponding
velocity field. The Supergalactic Plane can be clearly seen on the $SGZ=0\kms$
slice, with Centaurus and P-P visible as strong overdensities. The
velocity field on the same plane shows the tug-of-war between these
two superclusters, and the resultant effect on the local group. There has 
been some controversy over the presence of backside infall in the GA region
(e.g. Dressler \& Faber 1990, Mathewson, Ford \& Buchhorn 1992). 
In this reconstruction,
backside infall can clearly be seen for both Centaurus/GA (centred at
$SGX\simeq-3500$\kms, $SGY\simeq1500$\kms in figure 4e) and P-P 
(centred at $SGX\simeq5500$\kms, $SGY\simeq-1500$\kms in figure 4e).
A cone including the Centaurus/Great Attractor region 
($305\deg < l < 325\deg$, $10\deg < b < 25\deg$) 
is shown in \S 6 as figure 9c. This shows velocity turnaround at 
$r \simeq 3500\kms$, with line-of-sight infall velocity reaching a 
maximum of $\sim 300\kms$ at $r \simeq 4500\kms$.

\bigskip
\noindent {\bf 5. Effects of Parameters on Reconstruction}
\medskip

In order to assess the effects of the parameters discussed in \S 3,
the reconstruction was also carried out with a variety of different
parameter values. When expanding the initial, redshift-space density
field in terms of harmonics, the expansion at some radial and angular
mode. This determines the resolution of the initial expansion, and
hence the final reconstruction, as the whole procedure is carried out
in terms of these same harmonics. If the expansion is terminated at a
very low order, structures on a smaller scale than wavelength of the
highest harmonic will be lost. On the other hand, if the expansion is
continued to a very high order, we risk introducing artefacts as the
density field ripples on scales smaller than the real
structures. Appendix B of FLHLZ contains a detailed discussion of
these problems.

Figure 6 shows the density field on the $SGZ = 0\kms$ plane,
demonstrating how additional structure is resolved as angular modes
are added. In figure 6a, the angular modes are terminated at $l_{\rm max}
= 4$, and the circular structures associated with low-frequency
angular modes are very clear. Overdensities occur on distinct shells,
and follow circular paths about the origin; furthermore, resolution
falls with radius, as the wavelength of the highest order mode
increases. In figure 6b, $l_{\rm max} = 10$, and the structures are much
more detailed, with less obvious artefacts. The arc across the top
right still follows a circular path, and do the two clusters at the
bottom left. Nonetheless, structure is resolved quite well, with both
the P-P and Centaurus/GA regions clearly visible. Comparing these to
the reconstructions earlier in this paper, where $l_{\rm max} = 15$, it is
interesting to note that the purely angular arc across the top left of
the $SGZ = 0\kms$ slice is still present.

Beyond those parameters inherent in the expansion, 
a choice of cosmological
parameters is necessary for the actual reconstruction. The shape of
the prior power spectrum, characterised by $\Gamma$ and $\sigma_8$,
seems not to affect the reconstruction strongly within $r <
10,000\kms$. We have also performed the reconstruction of the IRAS
velocities with a standard CDM ($\Gamma=0.5$) prior. Since the
standard CDM model has less large scale power than the $\Gamma=0.2$
model, the WF smoothes more on large scales and therefore the
reconstructed velocities tend to be smaller; the overall difference is
however small with $\langle Delta_{\rm RMS} \rangle\simlt 50 \kms$.
Figure 7 shows density and velocity fields on the $SGZ = 0\kms$ plane
for two different prior power spectra; ($\Gamma=0.2$, $\sigma_8=1.0$)
and ($\Gamma=0.5$, $\sigma_8=0.7$). Although the choice of $\beta$ 
plays a complex role in the mechanics of the reconstruction, changes
within the range $0.2 < \beta < 1.0$ have little effect other than
a linear scaling of peculiar velocities.
Figure 8 shows the
density fields on the $SGZ = 0\kms$ plane for $\beta=0.5$
and $\beta=1.0$, and the residual differences in the density and 
velocity fields.
The residual velocity field is calculated by correcting for linear velocity
scaling such that $|{\bf v}| \propto \beta$. Hence:
$$
{\bf v}_{\rm resid} = {{{\bf v}_1}\over{\beta_1}} - {{{\bf v}_2}\over{\beta_2}}
\quad .
\eqno(7) 
$$
As can be seen from the plots, there is very little discrepancy from this
linear scaling, indicating that within this volume, the reconstruction
is robust to changes in $\beta$.

\bigskip
\noindent {\bf 6. Comparison with Mark III}
\medskip

The Mark III catalogue (Willick \etal 1995) 
contains about 3000 galaxy peculiar velocities from
Tully-Fisher distances. In figure 9, we show
comparisons between Mark III velocities and those reconstructed from
the IRAS 1.2Jy survey. These are plotted for cones selected by
Galactic latitude and longitude, following Faber \& Burstein (1988).
Throughout these plots, it is important to note that the Mark III data is
unsmoothed and hence displays considerably more scatter than that from
WF reconstruction. As such, it is not possible to make anything more than
a qualitative comparison. Davis, Nusser \& Willick (1996) recently 
made a mode-by-mode comparison of IRAS and Mark III using spherical
harmonics, finding a reduced $\chi^2 \sim 2$. A summary of other 
comparisons is given in Dekel (1994).

Figure 9a shows the central region of the Virgo cluster ($265\deg < l
< 315\deg$, $67\deg < b < 80\deg$). There is good agreement on the
velocity gradient at around 1500 \kms, where data are rich in both
catalogues. However, for $2000 \kms < r < 3000 \kms$, the WF
velocities are near zero, while Mark III shows a continuation of
the earlier velocity gradient, giving strong backside infall 
($v_{\rm radial} \simeq -1200\kms$ 
at $r \simeq 2500\kms$). A similar effect is seen in figure 9b,
which shows the Leo Cloud ($200\deg < l < 260\deg$, $50\deg < b <
70\deg$). Velocity gradients in the two catalogues are very
similar out to about 1500 \kms, after which the WF levels out while
the Mark III continues to show back-side infall out to $r \simeq 3000
\kms$; beyond this, Mark III data is dominated by scatter. 
The Centaurus/Great Attractor region ($305\deg < l < 325\deg$, $10\deg < 
b < 25\deg$) is shown in figure 9c. In both surveys clear back-side infall 
can be seen, with turn-around at $r \simeq 3500 \kms$. Apart from the
expected scatter in the Mark III, there is very good agreement. 
Figure 9d shows the Fornax-Eridanus region ($193\deg < l < 245\deg$, 
$-66\deg < b < -46\deg$). Again, the Mark III data shows considerably
higher velocity contrast than the WF reconstruction, with a $\sim 1000\kms$
velocity difference across the nearby cluster, compared to $\sim 300\kms$ 
for WF.

Many of these differences might well be inherent to the WF procedure,
which, by construction, goes to $v_{\rm pec} = 0$ at large scales. 
In regions of
sparse IRAS data, this could account for the reduced velocity contrast in
the WF reconstruction. In principle, a comparison between reconstructed
velocities and those from direct measurement could be used to constrain
power from larger scales.

\bigskip
\noindent{\bf 7. The Acceleration of the Local Group} 
\medskip

% {\it Figure 10 - Predicted dipole within radius R as a function of R}

% {\it Figure 11 - The direction of the LG dipole}

The dipole temperature anisotropy of the microwave background (CMB)
has been measured by the COBE satellite to extraordinary accuracy,
$D=3.343\pm0.016\, {\rm mK}$ in the direction $(l=264.4\deg\pm
0.3\deg,\ b=48.4\deg \pm 0.5\deg)$ (Smoot \etal 1991, 1992; Kogut
\etal 1993; Fixsen \etal 1994). Although various alternative theories
have been proposed (e.g., Gunn 1988; Paczy\'nski \& Piran 1990; 
Langlois \& Piran 1996), this
anisotropy is usually interpreted as due to the motion of the Earth
with respect to a rest frame defined by the CMB. After correcting the
Earth's motion relative to the Local Group (LG) barycenter, one infers
(Smoot \etal 1991, Kogut \etal 1993) that the LG is moving at a
velocity of $627\pm 22$ \kms\ in a direction $(l=276\deg\pm 3\deg,\
b=30\deg\pm 3\deg)$.

In the gravitational instability scenario, the peculiar velocity of
the LG is generated by surrounding fluctuations in the mass
density. In linear theory, the relation between the gravitational
acceleration induced by the mass inhomogeneities and the LG dipole is
particularly simple, 
$$ 
\bfv({\bf 0}) = {{f(\Omega)H_0}\over{4\pi}}\, \int d^3\bfr\, \delta(\bfr)\, {{\hat\bfr}\over{r^2}} \quad .
\eqno(8) 
$$ 
In the context of the linear biasing model adopted in our
reconstruction procedure, the the integral over the mass fluctuations
can be replaced by an integral over the galaxy fluctuations provided
we allow for a relative bias, i.e., $f(\Omega)\to \beta=
f(\Omega)/b$. Note the dipole is independent of the extragalactic
distance scale if we measure distances in \kms.

The integral in equation 8 can be evaluated using a galaxy redshift
survey. By comparing the result with the known LG velocity, one infers
an estimate of $\beta$. This technique (and variants using a
flux-weighted dipole) has been applied by many authors to both optical
and infrared galaxy samples as well as clusters catalogues (see the
review by Strauss \& Willick and references therein). The main
uncertainty in the value of $\beta$ determined in analyses of the
dipole stem from statistical noise due to sparse sampling, the effects
of redshift distortion, and the unknown contribution to the dipole
from scales larger than the sample size or structures hidden 
behind the Zone of Avoidance.

The dipole predicted from equation 8 is (with our prescription of
linear bias) linear in the galaxy density field. Consequently, the
dipole computed from our WF algorithm will be an optimal (in the sense
of minimum variance) estimate of the dipole due to matter within the
reconstruction volume to the extent that non-linear effects can be
neglected. Figure 10 shows the predicted dipole generated by matter
within radius $R$ as a function of $R$ as computed from our WF
reconstruction (see Appendix A) for our canonical prior
($\Gamma=0.2$, $\sigma_8=0.7$, and $\beta=0.7$). The amplitude of
reconstructed dipole for the model is slightly larger than the dipole
observed by COBE. The amplitude reconstructed dipole is roughly
proportional to $\beta$ (there is some non-linear dependence which
enters in the correction for the redshift distortion, see equations 16
and 18 of FLHLZ) and therefore a reconstruction with value of $\beta$
slightly smaller than our canonical model would provide a better
fit. The direction of the LG dipole is shown in figure 11. 
The misalignment angle between the IRAS and CMB Local
Group dipoles is only $13\deg$ out to $R=50$\mpc, 
partially due to the `tug-of-war' between the Centaurus/GA and 
P-P superclusters.
However, at $R=200$\mpc, 
the dipole of our canonical reconstruction points in the
direction $(l=247\deg,\ b=37\deg)$ which is $25\deg$ away from the
COBE direction.
As our procedure suppresses shot-noise this misalignment may 
indicate non-linear effects on these very large scales. However, 
assessing the
significance of this misalignment is difficult, as the dipole observed
from any volume-limited sample will inherently show scatter about
the 'true' CMB value (Vittorio \& Juszkiewicz 1987, Juszkiewicz, 
Vittorio \& Wyse 1990, Lahav, Kaiser \& Hoffman 1989). 
For example, Lahav \etal (1989) show that for
a biased CDM universe, 85\% of observers will see a misalignment angle
of $\theta < 20\deg$ in a sample out to $r = 4000\kms$.

In figure 11, we show the scatter in the reconstructed dipole due
statistical noise and from contributions from fluctuations outside our
reconstruction volume. The statistical scatter about the dipole can be
computed within the framework of the WF algorithm and the result for
the scatter in the dipole is given in Appendix B. This scatter is
shown in figure 11 as the set dotted curves; this scatter grows with
distance as the smoothing by the WF increases to mitigate the effects
of shot noise. The uncertainty in the dipole due to matter outside the
reconstruction volume can be calculated given the power spectrum
assumed in the prior; the result is given in Appendix B and shown (for
our canonical prior) as the dashed lines in figure 11. This scatter
decreases steadily to zero as $R$ increases. We should point out that
the dipole generated within $r<R$ is a well defined quantity.

\bigskip
\noindent{\bf 8. Bulk Velocities} 
\medskip

% {\it Figure 12 - Cartesian components of the bulk flow}

A simple statistic of the peculiar velocity field is the average or
bulk velocity within a window, $W(\bfr)$, 
$$ 
\langle \bfv\rangle_{R_s} = \int\limits_{r<R_s} d^3\bfr\, \bfv(\bfr)\, W(\bfr) \quad . 
\eqno(9) 
$$ 
A commonly used window is spherical top-hat, $W(r)
=(4\pi/3R^3)^{-1}\Theta(r-R)$, where $\Theta(x)$ is the usual step
function. The bulk velocity is sensitive to the fluctuations on scales
$>R$ and thus probes the linear regime even when non-linear effects
in the density field on scales $<R$ are important.

It is straightforward to relate the bulk velocity of spherical window
centred on the LG to the expansion of the density field in spherical
harmonics (see Appendix C). Once again, the WF reconstructed velocity
field can be used to compute an optimal estimate of the bulk velocity
in linear theory. As in the dipole calculation there will be scatter
introduced by dilute sampling and fluctuations outside the
reconstruction which can be computed for a given prior model of power
spectrum (see Appendix C).

In figure 12, we show the Cartesian components of the bulk flow
computed within spherical windows of radius, $R$. The points with
error bars show the WF reconstruction. 
The curves show the values from
the Potent analysis (Dekel 1994 and private communication) of 
the Mark III compilation of peculiar velocities (Willick \etal 1995).
The reconstructed IRAS bulk
flow out to $5000 \kms$ is $\sim 300 \kms$, which agrees in amplitude
with that derived from the Mark III peculiar velocities ($\sim 370
\kms$). However, the two vectors deviate by $\sim 70\deg$ in direction; 
there is a 33 percent probability of two random vectors aligning 
within $70\deg$.
This discrepancy could be due to differences in the width of the zone of 
avoidance and other systematic effects.

\bigskip
\noindent{\bf 9. The Extent of the Supergalactic Plane} 
\medskip

The so-called Supergalactic Plane (SGP) was recognised by de
Vaucouleurs (1956) using the Shapley-Ames catalogue. This followed 
earlier work by Vera Rubin, whose analysis of radial velocities of 
nearby galaxies suggested a differential rotation of the `metagalaxy'.
Indeed, the SGP had in fact already been noted
by William Herschel more than 200 years earlier.
Traditionally the Virgo cluster was regarded as the centre of the
Supergalaxy, and this was termed the `Local Supercluster'. 
However, recent
maps (e.g. figures 2-5) of the local universe indicate that much
larger clusters (e.g. Great Attractor, Perseus-Pisces) are major 
components of this `plane'.
The north pole of the standard SGP (de Vaucouleurs 1976) lies in
the direction of Galactic coordinates ($l=47.37\deg$, $b= 6.32\deg$). The
origin of SGL is at ($l=137.37\deg$, $b=0\deg$), which is one of the two
regions where the SGP is crossed by the Galctic Plane. The Virgo
cluster is at SGP coordinates ($SGL=104\deg$, $SGB = -2\deg$). 
 
Although the SGP is clearly visible in whole-sky galaxy catalogues,
it has only been re-examined quantitatively in recent
years. Tully (1986) claimed that the flattened distribution of clusters
extended across a diameter of $\sim 0.1 c$ with axial ratios of 4:2:1.
Shaver \& Pierre (1989) found that radio galaxies were more strongly
concentrated towards the SGP than were optical galaxies, 
and that the SGP (as
represented by radio galaxies) extended out to redshift $z \sim 0.02$. 
Di Nella \& Paturel (1995) examined the SGP using a
compilation of nearly 5700 galaxies larger than 1.6 arcmin, and found
qualitatively agreement with the standard SGP. Lahav \etal (1996,
in preparation; also Lahav 1996) revisited the
SGP using the Optical Redshift Survey (Santiago \etal 1995)
and the IRAS 1.2 Jy survey (Fisher \etal 1995) 
To objectively identify a `plane' they 
calculated the moment of inertia 
(with the observer located at the centre)
by direct summation over the galaxies (taking into account the 
selection function) and subtracting the mean background density 
$n_{\rm bg}$ in the absence of the `plane'.
By finding the eigenvalues and vectors of the inertia tensor
they deduced that the derived 'plane' is aligned to within 
$30\deg$
of the standard SGP.
The diameter of the SGP in ORS and IRAS 
was estimated to be at least 12,000 \kms.

Here we take a different approach as the reconstructed field
is continuous.
Again, we can use our formalism to obtain the optimal reconstruction
(in the minimum variance sense) of the moment of inertia. As
explained earlier, a convenient property of this Wiener approach applied to
the density fluctuation field, $\delta$, is that it will also give
the optimal reconstruction for any property which is linear in $\delta$. 
In particular, if we seek the optimal reconstruction of the
moment of inertia 
$$
{\tilde C_{ij}} = 
\; C_{ij} - {{\bar C}_{ij} } = 
\; {1 \over N}
\int\!\!\!\int\!\!\!\int \left[ \rho \left( {\bf x} \right) - n_{\rm bg} \right] 
\left( x_i - {\bar {x_i} } \right) \left( x_j
 - {\bar {x_j}} \right) 
\; dV
\qquad .
\eqno (10)
$$
This can be re-written as 
$$
{{\tilde C}_{ij} } = 
\left( { 3 \over { 4 \pi R^3} } \right) I_{ij} +
\left( \left[ 1 - { n_{\rm bg} \over { \bar \rho}} \right] { R^2 \over 5} \right)
\delta_{ij}^k
\qquad ,
\eqno (11)
$$
where $\delta^k$ is the Kroneker delta and
$$ 
I_{ij} = \int_R \delta ({\bf r} ) x_i x_j dV 
\qquad .
\eqno (12)
$$
This can be expressed {\it analytically} in terms of the reconstructed
coefficients $\delta_{lmn}^R$. The full mathematical details are
given in Appendix D. We emphasise again that in the
Wiener approach the density field goes to the mean density at large
distances. This does not necessarily mean that the SGP itself
disappears at large distances, it only reflects our ignorance of what
exists out there, where only very poor data are available. 

To find the alignment and extent of the 'plane', we diagonalise
the covariance matrix and find the eigen-values and eigen-vectors :
$$
{\tilde C} {\bf u}_\alpha = \lambda_\alpha {\bf u}_\alpha
\qquad ,
\eqno (13)
$$
where the $\lambda_\alpha$'s and ${\bf u}_\alpha$'s are the
eigen-values and eigen-vectors respectively ($\alpha=1,2,3$). 
The `half-width'
(1-sigma) along each of the 3 axes is given by
$\sqrt{\lambda_{\alpha}}$. Note that since the background
contribution (the last term in equation 11) is isotropic, it only affects
the eigen-values, but not the directions of the eigen-vectors. 

We have applied this technique to the reconstructed, real-space density
field, assuming as priors
$\beta=0.7$, $\sigma_8 = 0.7$ and a CDM power spectrum with
$\Gamma =0.2$.
Out to $R_{\rm max} = 4000, 6000, 8000 \kms$ 
the derived plane is aligned with de Vaucouleurs' SGP to within 
$15\deg$, $35\deg$ and $31\deg$ respectively. 
Note that the 
probability of 2 random vectors being aligned within an angle 
$\theta$ is
$$
P(< \theta) = {{1-\cos(\theta)}\over{2}}
\qquad .
\eqno (14)
$$
Hence, for $\theta =30\deg$ only $P \sim 7 \%$. 

Out to $R_{\rm max} = 4000 \kms$ 
the axial ratio is roughly $7:5:1$, assuming a background ratio 
$(n_{\rm bg} / {\bar\rho}) = 0.79$, derived from the IRAS data.
The results are only slightly different for raw harmonics, uncorrected 
for redshift distortion and noise, suggesting that redshift distortion 
is negligible on these large scales. 
The results also agree well with those direct 
summation of the moment of inertia (Lahav \etal 1996, in preparation).

\bigskip
\noindent{\bf 10. Conclusions \& Discussion} 
\medskip

In this paper, we have applied Wiener reconstruction with spherical 
harmonics and Bessel functions to the IRAS 1.2Jy redshift survey. Using
a prior based on parameters $\beta=0.7$, $\sigma_8 = 0.7$ and a CDM 
power spectrum with $\Gamma =0.2$, we find that:

\item{1.} The reconstructed density field clearly shows many known
structures, including N1600 (Santiago \etal 1995) and A3627
(Kraan-Korteweg \etal 1996). Notably, Perseus-Pisces is seen
to extend out to $R \sim 9000\kms$, while Virgo, Centaurus and 
Telescopium-Indus-Pavonis join to form a large structure extending
over $1500\kms < R < 7000\kms$. A number of new clusters are also
observed. Detailed maps are shown in \S 4.

\item{2.} The reconstructed velocity field shows backside infall 
for both Perseus-Pisces and the the Centaurus/Great Attractor region.
The Centaurus/Great Attractor supercluster shows velocity turnaround at 
$r \simeq 3500\kms$, with line-of-sight infall velocity reaching a 
maximum of $\sim 300\kms$ at $r \simeq 4500\kms$. Further,
there is reasonable qualitative agreement between the reconstructed 
velocity field and that derived from Tully-Fisher measurements (Mark III).

\item{3.} The misalignment angle between the CMB and reconstructed 
IRAS LG dipoles falls to a minimum of $13\deg$, calculated for
$R<50$\mpc, but increases to $25\deg$ for $R<200$\mpc.

\item{4.} The reconstructed IRAS bulk flow out to $5000 \kms$ is 
$\sim 300 \kms$, which agrees in amplitude with that derived from 
the Mark III peculiar velocities ($\sim 370\kms$). 
However, the two bulk flow vectors deviate by some $\sim 70\deg$.

\item{5.} The alignment and extent of the reconstructed Supergalactic 
Plane can be determined analytically from the harmonic coefficients.
Out to $R_{\rm max} = 4000, 6000, 8000 \kms$ the derived plane is 
aligned with de Vaucouleurs' SGP to within $15\deg$, $35\deg$ and 
$31\deg$ respectively. Out to $R_{\rm max} = 4000 \kms$ 
the axial ratio is roughly $7:5:1$.

\noindent We also confirm that the reconstruction is robust to changes 
in prior parameters.

Wiener reconstruction is particularly well suited to recovering 
real-space density and velocity fields from near whole-sky surveys.
The WF formalism provides a rigorous methodology for variable smoothing,
determined by the sparseness of data relative to the expected signal.
As such, natural continuations of this work would be application to
newer surveys, such as the PSCZ (Saunders \etal, in preparation), 
which is complete down to 0.6Jy, or ORS (Santiago \etal 1995).

\vfill\eject 

\bigskip
\noindent{\bf Acknowledgements} 
\medskip

We would like to thank Avishai Dekel, Yehuda Hoffman, Donald Lynden-Bell,
Basilio Santiago and Selim Zaroubi for their help in preparing this paper.
We are also grateful to Steven Gardner for his assistance in identifying
observed clusters.

\bigskip
\bigskip
\centerline{\bf References} 
\bigskip

\pp{Abell, G.O., Corwin, H.G., \& Olowin, R.P., 1989, ApJ Suppl., 70, 1}

\pp{Bunn, E., Fisher, K.B., Hoffman, Y., Lahav, O., Silk, J., \& Zaroubi, S., 1994, preprint}

\pp{Davis M., Nusser A. \& Willick J.A., ApJ in press, available as SISSA astro-ph/9604101}

\pp{Dekel, A. 1994, ARAA, 32, 371}

\pp{Dressler, A., \& Faber, S. M. 1990, ApJ, 354, L45}

\pp{Efstathiou, G., Bond, J.R., \& White, S.D.M., 1992, MNRAS, 258, P1}

\pp{Fabbri, R., \& Natale, V., 1989, ApJ, 363, 53}

\pp Faber, S.M. \& Burstein 1988, in {\it Large Scale Motions in the Universe: A Vatican Study Week}, eds. V.C. Rubin \& G.V. Coyne, S.J. (Princeton: Princeton Univ. Press), p. 115

\pp{Feldmann, H., Kaiser, N., \& Peacock, J., 1994, ApJ, 426, 23}

\pp{Fisher, K.B., Davis, M., Strauss, M.A., Yahil, A., \& Huchra, J.P., 1994$a$, MNRAS, 266, 50}

\pp{Fisher, K.B., Davis, M., Strauss, M.A., Yahil, A., \& Huchra, J.P., 1994$b$, MNRAS, 267, 927}

\pp{Fisher, K.B., Davis, M., Strauss, M.A., Yahil, A., \& Huchra, J.P., 1993, ApJ, 402, 42}

\pp{Fisher, K.B., Scharf, C.A. \& Lahav, O., 1994, MNRAS, 266, 219 (FSL)}

\pp{Fisher, K.B., Lahav, O., Hoffman, Y., Lynden-Bell, D., \& Zaroubi, S., 1995$a$, MNRAS, 272, 885 (FLHLZ)}

\pp{Fisher, K.B., Huchra, J.P., Strauss, M.A., Davis, M., Yahil A., Schlegel D., 1995$b$, ApJ, 100, 69}

\pp{Fixsen, D. E., Cheng, E. S., Cottingham, D. A., Eplee, R. E., \& Isaacman, R. B., \etal 1994, ApJ, 420, 445}

\pp{Gunn, J. E. 1988, in ASP Conf. Ser., Vol. 4, The Extragalactic Distance Scale, ed. S. van den Bergh \& C. J. Pritchet (San Francisco: ASP), 344}

\pp{Heavens, A.F., \& Taylor, A.N., 1995, MNRAS, 275, 483}

\pp{Juszkiewicz, R., Vittorio, N., \& Wyse, R.F.G., 1990, ApJ, 349, 408}

\pp{Kaiser, N. \& Stebbins, A., 1991, in {\it Large Scale Structure and Peculiar Motions in the Universe}, eds. D.W. Latham \& L.N. DaCosta (ASP Conference Series), p. 111}

\pp{Kaiser, N., 1987, MNRAS, 227, 1}

\pp{Kaiser, N., Efstathiou, G., Ellis, R., Frenk, C., Lawrence, A., Rowan-Robinson, M., \& Saunders, W., 1991, MNRAS, 252, 1}

\pp{Kogut, A., Lineweaver, C., Smoot, G. F., Bennett, C. L., Banday, A., \etal 1993, ApJ, 419, 1}

\pp{Kraan-Korteweg, R.C., Woudt, P.A., Cayatte, V., Fairall, A.P., Balkowski, C., \& Henning, P.A., 1996, Nature, 379, 519}

\pp{Lahav, O., 1992, in {\it Highlights of Astronomy}, vol. 9, p. 687, the XXIst General Assembly of the IAU, ed. Bergeron J., Kluwer, Dordrecht.}

\pp{Lahav, O., Yamada, T., Scharf, C.A. \& Kraan-Korteweg, R.C., 1993, MNRAS, 262, 711}

\pp{Lahav, O., Fisher, K.B., Hoffman, Y., Scharf, C.A, \& Zaroubi, S., 1994, ApJL, 423, L93 (LFHSZ)}

\pp{Lahav, O., 1996, in {\it Mapping, Measuring and Modelling the Universe}, Valencia September 1995, eds. P. Coles \etal, APS Conference Series}

\pp{Lahav, O., Santiago, B.X., Strauss, M.A., Webster, A.M., Davis, M., Dressler, A. \& Huchra, J.P., 1996, in preparation}

\pp{Langlois, D., \& Piran, T., 1996, Phys. Rev. D., 53, 2908} 

\pp{Lynden-Bell, D. 1991., in {\it Statistical Challenges in Modern Astronomy}, eds. Babu G.B. \& Feigelson E.D.}

\pp{Mathewson, D.S., Ford, V.L., \& Buchhorn, M., 1992, ApJ, 389, L5}

\pp{Nusser, A., \& Davis, M., 1994, ApJL, 421, L1 (ND)}

\pp{Paczy\'nski, B., \& Piran, T. 1990, ApJ, 364, 341}

\pp{Peebles, P.J.E., 1973, ApJ, 185, 413}

\pp{Pellegrini, P.S, Dacosta, L.N., Huchra, J.P., Latham, D.W., \& Willmer, C.N.A., 1990, Astron. J, 99, 751}

\pp{Press, W.H., Teukolsky, S.A., Vetterling, W.T., \& Flannery, B.P.1992, Numerical Recipes (Second Edition) (Cambridge: Cambridge University Press)}

\pp{Reg\H{o}s, E. \& Szalay, A.S., 1989, ApJ, 345, 627}

\pp{Rybicki, G.B., \& Press, W.H., 1992, ApJ, 398, 169}

\pp{Santiago, B.X., Strauss, M.A., Lahav, O., Davis, M., Dressler, A., \& Huchra, J.P., 1995, ApJ, 446, 457}

\pp{Saunders, W. \etal, 1991, Nature, 342, 32}

\pp{Scharf, C., Hoffman, Y., Lahav, O., \& Lynden-Bell, D., 1992, MNRAS, 256, 229}

\pp{Scharf, C.A. \& Lahav, O., 1993, MNRAS, 264, 439}

\pp{Smoot, G. F., Bennett, C. L., Kogut, A., Aymon, J., Backus, C., \etal 1991, ApJL, 371, L1}

\pp{Smoot, G. F., Bennett, C. L., Kogut, A., Wright, E. L., Aymon, J., \etal 1992, ApJL, 396, L1}

\pp{Strauss, M.A. \& Davis, M., 1988, in {\it Large Scale Motions in the Universe: A Vatican Study Week}, eds. V.C. Rubin\& G.V. Coyne, S.J. (Princeton: Princeton Univ. Press), p. 256 }

\pp{Strauss, M.A., \& Willick, J.A., 1995, Phys Rev, 261, 271}

\pp{Tully, R.B., 1987, ApJ, 321, 280}

\pp{Vittorio, N., \& Juszkiewicz, R., 1987, ApJ, 314, L29}

\pp{Wiener, N., 1949, in {\it Extrapolation and Smoothing of Stationary Time Series}, (New York: Wiley)}

\pp{Willick, J. A., Courteau, S., Faber, S. M., Burstein, D., \& Dekel, A. 1995, ApJ, 446, 12 }

\pp{Yahil, A., 1988, in {\it Large Scale Motions in the Universe: A Vatican Study Week}, eds. V.C. Rubin\& G.V. Coyne, S.J. (Princeton: Princeton Univ. Press), 219}

\pp{Yahil, A., Strauss, M.A., Davis, M., \& Huchra, J.P., 1991, ApJ, 372, 380 (YSDH)}

\pp{Zaroubi, S., Hoffman, Y., Fisher, K.B., \& Lahav, O., 1995, ApJ, 449, 446}

--------------------------------------------------------------------------

\vfill\eject 
\centerline{\bf Appendices} 
\bigskip

\bigskip
\noindent{\bf Appendix A: Wiener Filtering } 
\medskip

Here, we give a brief review of the Wiener Filter technique; the
reader is referred to LFHSZ, ZHFL, and Rybicki \& Press (1992) for
further details. Let us assume that we have a set of measurements,
$\{d_\alpha\}\ (\alpha=1,2,\dots N)$ which are a linear convolution of
the true underlying signal, $s_\alpha$, plus a contribution from
statistical noise, $\epsilon_\beta$, such that
$$
d_\alpha = {\cal R}_{\alpha\beta}\left [ s_\beta +
\epsilon_\beta\right]
\qquad ,
\eqno(A1) 
$$
where ${\cal R}_{\alpha\beta}$ is the response or ``point spread''
function (summation convention assumed). In the present context it 
would be the radial coupling matrix
discussed in the previous section. Notice that we have assumed that
the statistical noise is present in the underlying field and therefore
is convolved by the response function.

The WF is the {\it linear} combination of the observed data which is
closest to the true signal in a minimum variance sense. More
explicitly, the WF estimate is given by $s_\alpha (WF) =
F_{\alpha\beta}\, d_\beta$ where the filter is chosen to minimise
$\langle |s_\alpha(WF)-s_\alpha|^2\rangle$. It is straightforward to
show (see ZHFL for details) that the WF is given by 
$$
F_{\alpha\beta} = \langle s_\alpha d_\gamma \rangle
\langle d_\gamma d_\beta^\dagger \rangle^{-1}\qquad ,
\eqno(A2) 
$$
where
$$
\langle s_\alpha d_\beta^\dagger\rangle = {\cal R}_{\beta\gamma}\,
\langle s_\alpha s^\dagger_\gamma\rangle 
\eqno(A3) 
$$
$$
\langle d_\alpha d_\beta^\dagger\rangle = 
{\cal R}_{\alpha\gamma}\, {\cal R}_{\beta\delta}\, \left[ 
\langle s_\gamma s^\dagger_\delta\rangle
+ \langle \epsilon_\gamma\epsilon^\dagger_\delta\rangle\right] \qquad .
\eqno(A4) 
$$
In the above equations, we have assumed that the signal and noise are
uncorrelated. From equation A4, it is clear that in order to implement
the WF one must construct a {\it prior} which depends on the variance
of the signal and noise.

The dependence of the WF on the prior can be made clear by defining
signal and noise matrices given by $S_{\alpha\beta}=\langle s_\alpha
s^\dagger_\beta\rangle$ and $N_{\alpha\beta}=\langle \epsilon_\alpha
\epsilon^\dagger_\beta\rangle$. With this notation, we can rewrite
equation A4 as
$$
{\bf s}(WF) = {\bf S} \left[ {\bf S} + {\bf N}\right]^{-1} {\bf {\cal
R}}^{-1} {\bf d} \qquad .
\eqno(A5) 
$$
Formulated in this way, we see that the purpose of the WF is to
attenuate the contribution of low signal to noise ratio data and
therefore regularize the inversion of the response function. The
derivation of the WF given above follows from the sole requirement of
minimum variance and requires only a model for the variance of the
signal and noise. The WF can also be derived using the laws of
conditional probability if the underlying distribution functions for
the signal and noise are assumed to be Gaussian; in this more
restrictive case, the WF estimate is, in addition to being the minimum
variance estimate, also both the maximum a {\it posterior} estimate
and the mean field (see LFHSZ, ZHFL). For Gaussian fields, the mean
WF field can be supplemented with a realisation of the expected
scatter about the mean field to create a realisation of the field;
this is the heart of the ``constrained realisation'' approach
described in Hoffman \& Ribak (1991; see also ZHFL). 

As Rybicki \& Press (1992) point out, the WF is in general a biased
estimator of the mean field unless the field has zero mean; this is
not a problem here since we will perform the filtering on the density
fluctuation field which has, by construction, zero mean.

\bigskip
\noindent{\bf Appendix B: Dipole Velocity in Spherical Harmonics} 
\medskip

The velocity of the central observer in the CMB frame, commonly
referred to a ``dipole'', due to the fluctuations with $r<R$ can be
related to the spherical harmonic expansion of the density field
(see FLHLZ equation~C9), 
$$
\eqalignno{
\bfv({\bf 0})_{r<R}
& = {{\Omega^{0.6}}\over{4\pi}}\, \int\limits_{V_R} d^3\bfrp\,
\delta(\bfrp)\, {{\bfrp}\over{r^{\prime 3}}} & (B1) \cr
&= {{\beta}\over{4\pi}}\, \sum\limits_{lmn} C_{ln}
{{\delta^\rr_{lmn}}\over{k_n}}
\int\limits_0^R dr^\prime\, j_l(k_nr^\prime)\,
\int\limits_{4\pi} d\Omega\,
\hat\bfrp\, Y_{lm}(\hat\bfrp)
\cr
&= {{\beta}\over{4\pi}}\,\left( {{4\pi}\over{3}}\right)^{1/2}\,
\sum\limits_{n}\, {{C_{1n}}\over{k_n}}\,
\left(
-\sqrt{2} Re[\delta^\rr_{11n}]\, \hat{\bf x} +\sqrt{2}
Im[\delta^\rr_{11n}]\, \hat{\bf y} + Re[\delta^\rr_{10n}]\, \hat{\bf
z}\right)\,
\left( 1 - j_0(k_nR)\right)
%\int\limits_0^R dr\, j_1(k_nr)
\ . \cr}
$$
Here $Re[a]$ and $Im[a]$ refer to the real and imaginary parts of a
complex number, $a$. The last line provides a convenient expression
for the Cartesian components of the dipole. In the limit that
$R\to\infty$, we recover the true dipole, apart from statistical noise
and systematic errors introduced by non-linear effects. >From the last
line equation~(B1), we see that the dipole within $r<R$ can thus be
regarded as the sum of two terms: the true CMB dipole and a correction
for finite sample size (the $j_0(k_nR)$ term). In fact the latter term
involving the $j_0(k_nR)$ can be shown to be the velocity of a shell
at a distance $R$ in the CMB frame. 

There are two main sources of error in the dipole derived via
equation~(B1). First, there is a statistical noise introduced by the
finite sampling of the density field by the IRAS catalogue. The scatter
in the reconstructed dipole can be easily calculated in the framework
of the Wiener filter (see FLHLZ Appendix F for the scatter in the
reconstructed density and velocity field). The scatter in each
component of the dipole is given by 
$$
\langle \Delta\bfv({\bf 0})\rangle^2_{\rm{wf}}= {{\beta^2}
\over{12\pi}}\, \sum\limits_{n\np}\, 
C_{1n}\, C_{1\np}\, {{\left(1-j_0(k_n r)\right)\left(1-j_0(k_\np
r)\right)}
\over{k_n\, k_\np}} 
\left[\left( {\bf I} -{\bf F}_1 \right)\, {\bf S}_1\right]_{n\np}\,
\qquad ,\eqno(B2)
$$
where ${\bf F}_l = \bfSl (\bfSl+\bfNl)^{-1}$ is defined in terms of
the signal, $\bfSl$, and noise, $\bfNl$ matrices.

The second source of error is from fluctuations outside the volume
used in the Wiener reconstruction. The contribution to the dipole from
$r>R$, 
$$
\bfv_{out}({\bf 0}) = {{\beta}\over{4\pi}}\,
\int\limits_{r<R} d^3\bfrp\, \delta(\bfrp) {{\bfrp}\over{r^{\prime 3}}}\quad ,
\eqno(B3)
$$
introduces a 1-D rms scatter in the dipole
$$
\langle \Delta\bfv({\bf 0})\rangle^2_{out}= 
{{1}\over{3}}\, {{\beta^2}\over{2\pi^2}}\, \int dk\, P(k) 
\left[ j_o(kR)\right]^2\quad .
\eqno(B4)
$$
In the above equation, $P(k)$ is the linear power spectrum defined by
the prior. The total error in the reconstructed dipole is taken to be
the quadrature sum of equations B3 and B4.

\bigskip
\noindent{\bf Appendix C: Bulk Velocity in Spherical Harmonics} 
\medskip

The average or ``bulk'' velocity measured within a spherically
symmetric region defined by window function with characteristic scale
$r=R_s$ is defined by 
$$ 
\langle \bfv\rangle_{R_s} =
\int d^3\bfr\, W(r)\, \bfv(\bfr)
 \quad , \eqno(C1)
$$
where the volume integral of the window is taken to be unity. If we
rewrite the above expression in terms of the Fourier components of
$\bfv$, 
$$ 
\langle \bfv\rangle_{R_s} =
{{1}\over{(2\pi)^3}}
\int d^3\bfk\, \bfv_\bfk \, W(kR_s) \qquad , 
$$
where $W(kR_s)$ is the Fourier transform of the window function
($W(kR_s)=3j_1(kR_s)/kR_s$ for spherical top-hat). From the above
equation, we see that the bulk velocity for region centred on the LG
is formally equivalent to the dipole velocity of LG induced by the
density field obtained from smoothing with the window. Consequently,
the analogous formulae for the bulk velocity, uncertainty due to
sampling, and the contributions from fluctuations outside the
reconstruction volume in terms of the spherical harmonics are readily
derived by substituting 
$$
\delta_{lmn} \to 
\delta_{lmn}W(k_nR_s)\quad {\rm and}\quad 
P(k) \to P(k)W(k_nR_s)
$$ 
in equations~$B1$, $B2$, and $B4$

\bigskip
\noindent{\bf Appendix D: Moment of Inertia in Spherical Harmonics} 
\medskip
\bigskip
\bigskip

Consider the moment of inertia tensor within a sphere of radius 
$R$ around the origin:
$$
I_{ij} = \int_R \delta ({\bf r} ) x_i x_j d V \; .
\eqno (1) 
$$
We now express the density fluctuation in terms of spherical 
harmonics and Bessel functions:
$$
\delta({\bf r} ) =
\sum_l^{l_{\rm max}} \; \sum_{m=-l}^{m=+l} \; \sum_{n=1}^{n_{\rm max}(l)} C_{ln}\;
\delta_{lmn} \;\; j_l(k_{n} r) \;\; Y_{lm}({\bf \hat r}) \;, 
\eqno (2) 
$$
By substituting eq. (2) into eq. (1) and utilising the properties of the
spherical harmonics we find after some algebra 
the elements of the symmetric matrix $I_{ij}$:
$$
\eqalign{
I_{xx} &= \sum_{n=1}^{n_{\rm max}(2)}
   C_{2,n} B_{2,n} 
   \left[{
   \sqrt{{2\pi}\over{15}} 
   \big(\delta_{2,-2,n} + \delta_{2,2,n}\big) -
   \sqrt{{4\pi}\over{45}}\,\,
   \delta_{2,0,n}
   }\right]
   + {{\sqrt{4\pi}}\over{3}}
   \sum_{n=1}^{n_{\rm max}(0)}
   C_{0,n}B_{0,n}\delta_{0,0,n}
\cr
\cr
I_{yy} &= \sum_{n=1}^{n_{\rm max}(2)}
   C_{2,n} B_{2,n} 
   \left[{
   -\sqrt{{2\pi}\over{15}} 
   \big(\delta_{2,-2,n} + \delta_{2,2,n}\big) -
   \sqrt{{4\pi}\over{45}}\,\,
   \delta_{2,0,n}
   }\right]
   + {{\sqrt{4\pi}}\over{3}}
   \sum_{n=1}^{n_{\rm max}(0)}
   C_{0,n}B_{0,n}\delta_{0,0,n}
\cr
\cr
I_{zz} &= {{\sqrt{4\pi}}\over{3}}
   \left[{
   {{2}\over{\sqrt{5}}}
   \left({
   \sum_{n=1}^{n_{\rm max}(2)}
   C_{2,n} B_{2,n} \delta_{2,0,n}
   }\right)
   +
   \left({
   \sum_{n=1}^{n_{\rm max}(0)}
   C_{0,n} B_{0,n} \delta_{0,0,n}
   }\right)
   }\right]
\cr
\cr
I_{xy} &= -i{\sqrt{{2\pi}\over{15}}}
   \sum_{n=1}^{n_{\rm max}(2)}
   C_{2,n} B_{2,n} 
   \big(\delta_{2,-2,n} - \delta_{2,2,n}\big)
\cr
\cr
I_{xz} &= {\sqrt{{2\pi}\over{15}}}
   \sum_{n=1}^{n_{\rm max}(2)}
   C_{2,n} B_{2,n} 
   \big(\delta_{2,-1,n} - \delta_{2,1,n}\big)
\cr
\cr
I_{yz} &= -i{\sqrt{{2\pi}\over{15}}}
   \sum_{n=1}^{n_{\rm max}(2)}
   C_{2,n} B_{2,n} 
   \big(\delta_{2,-1,n} + \delta_{2,1,n}\big)
\cr
}
$$
where $n_{\rm max}(l)$ is the $n_{\rm max}$ corresponding to a 
particular value of $l$, and
$$
B_{ln} \equiv \int_0^R j_l(k_nr) r^4 dr
\qquad .
$$
Note that as the $I_{ij}$ is quadratic in the coordinates, 
the only harmonics to appear are of $l=0$ and $l=2$.

% --------------------------------------------------------------------------

% --------------------------------------------------------------------------

\vfill\eject 
\centerline{\bf Figure Captions} 
\bigskip 

Fig. 1 --- Scatter between velocities generated in Local Group and
CMB frames, plotted as a function of $\beta$ used in the 
reconstruction. The amplitude of the scatter is divided by $\beta$
so as to remove the linear scaling of velocities with $\beta$ 
inherent in reconstructions based on linear theory. A clear minimum
can be seen at $\beta \simeq 0.7$.

Fig. 2 --- The reconstructed density field, evaluated on thin shells 
at various real space distances, shown in Galactic Aitoff projection. 
Dashed lines show $\delta < 0$, and solid lines show $\delta \ge 0$, 
with contour spacing $\Delta \delta = 0.1$. (a) $2000\kms$: Clusters 
marked are N5864, Virgo (Vir), Centaurus (Cen), Hydra (Hyd), Fornax- 
Doradus-Eridanus (FDE), Ursa Major (Urs), and C$\alpha$. Voids marked 
are Local Void (LV), V$\alpha$ and V$\beta$. (b) $4000\kms$: Clusters 
marked are Centaurus (Cen), Telescopium-Indus-Pavonis (TIP), Hydra 
(Hyd), Leo, Cancer (Can), N1600, Camelopardalis (Cam), Perseus-Pisces 
(P-P), Cetus (Cet), C$\beta$, C$\gamma$ and C$\delta$. Voids shown 
are Local Void (LV) and V$\beta$. (c) $6000\kms$: A3627, Hydra (Hyd), 
Leo, Cancer (Can), Orion (Ori), A569, Perseus-Pisces (P-P), Pegasus 
(Peg), Telescopium-Indus-Pavonis (TIP), C$\gamma$, C$\delta$, 
C$\epsilon$ and C$\zeta$. One void is marked V$\gamma$. (d) 
$8000\kms$: Clusters marked are Great Wall (GW), A3627, Hydra (Hyd), 
Leo, A539, A779, A400, Coma (Com), Perseus-Pisces (P-P), Pegasus 
(Peg), Cygnus (Cyg), C$\gamma$, C$\zeta$ and C$\eta$. The 
continuation of V$\gamma$ is shown. 

Fig. 3 --- The reconstructed velocity field, evaluated on thin shells 
at various real space distances, shown in Galactic Aitoff projection.
These velocity fields correspond with the density fields shown in
figure 2. Dashed lines show infall, and solid lines show outflow. The 
first solid line is for $v_{\rm radial} = 0\kms$, and contour spacing 
is $\Delta v_{\rm radial} = 50\kms$. (a) $2000\kms$: Note the velocity 
dipole due to the Centaurus/GA region (towards $l \simeq 300\deg$, 
$b \simeq 15\deg$). (b) $4000\kms$: Note outflow towards P-P 
($l \simeq 145\deg$, $b\simeq -20\deg$), N1600 ($l \simeq 195\deg$, 
$b \simeq -20\deg$) and Cancer ($l \simeq 190\deg$, $b \simeq 20\deg$).
(c) $6000\kms$: Dominated by motion from the south Galactic pole 
towards the North Galactic pole, caused by outflow towards the Great 
Wall in the northern Galactic hemisphere. (d) $8000\kms$: Outflow
caused by large superclusters at $l \simeq 60\deg$, $b \simeq 60\deg$ 
(forming behind the Great Wall), and $l \simeq 220\deg$, 
$b \simeq -30\deg$ (forming behind C$\eta$ and A400). Also, note
strong infall can be seen behind Pegasus ($l \simeq 90\deg$, 
$b \simeq -25\deg$) and Hydra ($l \simeq 285\deg$, $b \simeq 5\deg$).

Fig. 4 --- The reconstructed density field on thin slices
at Supergalactic $x, y, z = 0, \pm 4000\kms$. Contours start
at $\delta = 0$ with spacing $\Delta \delta = 0.25$. Note the clear
presence of the Supergalactic plane in the $SGZ = 0\kms$ plot (e).

Fig. 5 --- The reconstructed velocity field on thin slices
at Supergalactic $x, y, z = 0, \pm 4000\kms$. Arrows show the
projection of the local peculiar velocity onto the plane, with length
giving the amplitude in accordance with the axis scale. These velocity
fields correspond to the density fields shown in figure 4. Note clear
backside infall towards both Centaurus/GA (centred at 
$SGX\simeq-3500\kms$, $SGY\simeq1500\kms$ in (e)) and P-P (centred at 
$SGX\simeq5500\kms$, $SGY\simeq-1500\kms$ in (e)).

Fig. 6 --- Cone diagrams comparing the IRAS reconstructed velocity
field with that derived from Tully-Fisher measurements (Mark III).
Hollow triangles show the IRAS reconstruction, while filled squares
show Mark III. (a) Central region of the Virgo cluster 
($265\deg < l < 315\deg$, $67\deg < b < 80\deg$). (b) Leo Cloud 
($200\deg < l < 260\deg$, $50\deg < b < 70\deg$). 
(c) Centaurus/Great Attractor region ($305\deg < l < 325\deg$, 
$10\deg < b < 25\deg$). (d) Fornax-Eridanus region 
($193\deg < l < 245\deg$, $-66\deg < b < -46\deg$). Note that the 
Mark III data is unsmoothed and hence displays considerably more 
scatter than that from WF reconstruction.

Fig. 7 --- Demonstration of different reconstruction resolutions.
The reconstructed density field, evaluated on a thin slice at 
$SGZ = 0\kms$. Contours start at $\delta = 0$ with spacing 
$\Delta \delta = 0.1$. Figure 5e shows the corresponding plot
for $l_{\rm max} = 15$; the default used throughout this paper.
(a) $l_{\rm max} = 4$: Clear circular patterns can be seen in the
density field. (b) $l_{\rm max} = 10$: Only a few circular artefacts
remain, given the addition of higher-order harmonics.

Fig. 8 --- Effects of varying cosmological parameters $\Gamma$ and 
$\sigma_8$. (a) and (c) show the reconstructed density field, 
evaluated for a thin slice at $SGZ = 0\kms$. Contours start at 
$\delta = 0$ with spacing $\Delta \delta = 0.1$. (b) and (d) show the 
reconstructed velocity field, evaluated for the same thin slice. 
Arrows show the projection of the local peculiar velocity onto the 
plane, with length giving the amplitude in accordance with the axis 
scale. Figures 5e and 6e show the corresponding plots for $\beta = 
0.7$, $\Gamma = 0.2$, $\sigma_8 = 0.7$; the canonical parameters used 
throughout this paper. (a) and (c) $\beta = 0.7$, $\Gamma = 0.5$, 
$\sigma_8 = 0.7$. (b) and (d) $\beta = 0.7$, $\Gamma = 0.2$, 
$\sigma_8 = 1.0$. 

Fig. 9 --- Effects of varying the cosmological parameter $\beta$. 
(a), (b) and (c) are density fields, evaluated for a thin slice at 
$SGZ = 0\kms$. Contours start at $\delta = 0$ with spacing $\Delta 
\delta = 0.1$. (a) and (b) show reconstructed density fields for 
$\beta$ of $0.5$ and $1.0$ respectively, with $\Gamma = 0.2$ and 
$\sigma_8 = 0.7$. Figures 5e shows the corresponding plot for $\beta 
= 0.7$, $\Gamma = 0.2$, $\sigma_8 = 0.7$; the canonical parameters 
used throughout this paper. (c) shows the residual density field 
after subtraction of (a) from (b), demonstrating the extremely small 
differences caused by change of $\beta$. Finally, (d) shows the 
residual velocity field, evaluated for the same thin slice. Arrows 
show the projection of the local peculiar velocity onto the plane, 
with length giving the amplitude in accordance with the axis scale.

Fig. 10 --- The amplitude of the acceleration, or dipole motion, of
the LG caused by the fluctuations within a sphere of radius $R$ versus
$R$. The heavy solid curve shows the dipole derived from the Wiener
reconstruction with our canonical prior ($\Gamma=0.2$, $\sigma_8=0.7$,
and $\beta=0.7$). The dotted curves show the expected scatter from
finite sampling (see Appendix B) while the dashed curves show the
scatter due to fluctuations outside the reconstruction volume (i.e.,
$r<200$\mpc); the dot-dashed curve is the quadrature sum of both
terms. The dotted horizontal line is the value of the LG dipole
inferred by COBE, $627\kms$. \bigskip

Fig. 11 --- The direction of the LG dipole. The crosses show the
convergence of the direction of the reconstructed dipole; starting at
the top of the plot the crosses give the direction (in Galactic
coordinates) of the dipole within $R$ (see Fig. 1) as $R$ is increased
in 1\mpc intervals. The direction of the dipole inferred from the
COBE measurements, $(l=276\deg,\ b=30\deg$; Smoot \etal 1991) is
denoted by the star. The circular curves denote angular separations
from the COBE result in $10\deg$ increments. \bigskip

Fig. 12 --- The average or bulk velocity within a sphere of radius $R$
centered on the Local Group a function of $R$. The four panels show
the three Cartesian components and the scalar amplitude of the bulk
flow. The points with error bars represent the bulk flows in the
Wiener reconstruction with our canonical prior ($\Gamma=0.2$,
$\sigma_8=0.7$, and $\beta=0.7$). The error bars are the scatter due
to both the finite sampling and fluctuations outside the
reconstruction volume (see Appendix C). The triangles connected by
the curves represent the measurements from the POTENT reconstruction
algorithm applied the Mark III peculiar velocity data (taken from
Dekel 1994). The two curves represent two different weighting schemes
and reflect the systematic uncertainty; the estimated random error is
approximately 15\% (Dekel 1994).

\end